\documentclass[usenatbib]{mn2e}
\usepackage{graphicx}
\usepackage{txfonts}

\title[$s$-process elements in young associations]{The chemical composition of nearby young associations:\\ $s$-process element abundances in AB Doradus, Carina-Near, and Ursa Major}
\author[V. D'Orazi et al.]{V. D'Orazi$^{1}$\thanks{E-mail: valentina.dorazi@mq.edu.au}, K. Biazzo$^{2}$, S. Desidera$^{3}$, E. Covino$^{2}$, S.M. Andrievsky$^{4,5}$, R.G. Gratton$^3$ \\
$^{1}$Department of Physics \& Astronomy, Macquarie University, Balaclava Rd., North Ryde, Sydney, NSW 2109, Australia\\
$^{2}$INAF Osservatorio Astronomico di Capodimonte, salita Moiariello 16, Napoli, 80131, Italy\\
$^{3}$INAF Osservatorio Astronomico di Padova, vicolo dell'Osservatorio 5, Padova, 35122, Italy\\
$^{4}$Department of Astronomy and Astronomical Observatory, Odessa National University,and Isaac Newton Institute of Chile,\\ Odessa Branch, Shevchenko Park, 65014 Odessa, Ukraine\\
$^{5}$GEPI, Observatoire de Paris-Meudon, CNRS, Universit\'{e} Paris Diderot, Meudon Cedex, 92125, France}
\begin{document}

\date{Accepted 2012 April 11. Received 2012 April 4; in original form 2012
February 29}

\pagerange{\pageref{firstpage}--\pageref{lastpage}} \pubyear{2012}

\maketitle

\label{firstpage}

\begin{abstract}
Recently, several studies have shown that young, open clusters 
are characterised by a considerable over-abundance in their barium content.
In particular, D'Orazi et al. (2009) reported that in some younger clusters [Ba/Fe] can reach values as high as $\sim$0.6 dex. The work also identified the presence of an anti-correlation 
between [Ba/Fe] and cluster age. For clusters in the age range $\sim$4.5 Gyr$-$500 Myr, this is best explained  by assuming a higher contribution from low-mass asymptotic giant branch stars to the Galactic chemical enrichment. 

The purpose of this work is to investigate the ubiquity of the barium over-abundance in young stellar clusters. 
We analysed high-resolution spectroscopic data, focusing on the $s$-process elemental 
abundance for three nearby young associations, i.e. AB Doradus, Carina-Near, and Ursa Major. 
The clusters have been chosen such that their age spread would complement the D'Orazi et al. (2009) study. 

We find that while the $s$-process elements Y, Zr, La, and Ce exhibit solar ratios 
in all three associations, Ba is over-abundant by $\sim$0.2 dex. 
Current theoretical models can not reproduce this abundance pattern, thus 
we investigate whether this unusually large Ba content might be related to chromospheric effects. 
Although no correlation between [Ba/Fe] and several activity indicators seems to be present, we conclude that different effects could be at work which may (directly or indirectly) be related to the presence of hot stellar chromospheres.

\end{abstract}


\section{Introduction}\label{sec:intro}
Elements heavier than iron cannot be produced by fusion reactions. Since the binding energy per nucleon decreases past the iron-group nuclei and because the Coulomb barrier increases, nuclei heavier than Fe are essentially synthesised by neutron ($n$) capture reactions. 
Two distinct processes are responsible for these heavier nuclei; the slow ($s$) neutron capture process during stellar He burning, 
and the rapid ($r$) neutron capture process\footnote{Neutron -capture reactions on iron seed can be slow and rapid relative to the average $\beta$-decay timescale of the radioactive nuclides.}, 
the latter being presumably related to Supernovae explosions (\citealt{ross}; \citealt{wana}). 
The $s$-process nucleosynthesis, which 
accounts for about half of the elemental abundances between Fe and Bi (see \citealt{busso99}; \citealt{kap11} for extensive reviews), is in turn produced through two channels:

$(i)$ the $weak$ component, which takes place in massive stars ($\gtrsim$8 M$_\odot$) during their advanced evolutionary stages of He-core or C-shell burning 
(e.g., \citealt{prant90}; \citealt{raite93}), is responsible in the Solar System 
for the production of the lightest isotopes (atomic mass number A$\lesssim$90; e.g., \citealt{tr04}).

$(ii)$ The $main$ $s$-process component is ascribed to thermally pulsing low-mass (M$\lesssim$ 4M$_\odot$) asymptotic giant branch (AGB) stars and accounts for the heaviest $s$-process elements, from Ba to Bi (e.g., 
\citealt{busso99}).

Since the abundance pattern of each component is dependent on a complex web of nuclear reactions, neutron fluxes, stellar masses and temperature ranges, $s$-process elements 
have long since been recognised as key diagnostics of the internal structure and the mixing phenomena in AGB stars. Furthermore, they  provide excellent tracers of 
the chemical enrichment mechanisms in the Galaxy and beyond.
For this reason, a wealth of observational studies have been performed over the years to derive $s$-process element abundances in both
field stars (halo, thick and thin disk, e.g., \citealt{mw95}; \citealt{mg01}; \citealt{bur})
and stellar clusters (globular and open clusters, \citealt{james04}; \citealt{yong08}; \citealt{mikolai}; \citealt{ele10}).

D'Orazi et al. (2009, hereafter D09) derived Ba abundances for a sample of 20 open clusters (OCs) spanning a wide range in age 
($\sim$30 Myr - 8 Gyr), Galactocentric distance (7-22 kpc), and metallicity ($-$0.3$<$[Fe/H]$<$+0.4 dex). 
They found that the [Ba/Fe] ratio increases from a roughly solar value ([Ba/Fe]$\sim$0 dex) for clusters 
with age $\gtrsim$ 4 Gyr, up to $\sim$0.2$-$0.3 dex for $\sim$500 Myr clusters . 
The anti-correlation between the Ba content and the cluster age can be reproduced by a Galactic chemical evolution model only assuming a higher Ba yield from 
low-mass AGB stars (i.e. 1$-$1.5 M$_\odot$) than that previously predicted (see D09 for further details). 
The same result was also confirmed by Maiorca et al. (2011, M11) who analysed both light $s$-process ($ls$; Y and Zr)
and heavy $s$-process elements ($hs$; La, Ce), concluding that the Ba enhancement is accompanied by similar trends from the other $s$-process elements.

Remarkably, when dealing with OCs younger than 500 Myr, D09 detected a further increase in [Ba/Fe]:
the three younger OCs of their sample, namely NGC~2516 ($\sim$110 Myr), IC~2391 ($\sim$50 Myr), and IC~2602 ($\sim$35 Myr) exhibit a
[Ba/Fe] ratio of 0.41$\pm$0.04, 0.68$\pm$0.07, and 0.64$\pm$0.07 dex, respectively. 
While a chemical evolution model with enhanced Ba production can account for the observed raising trend up to 
$\sim$500 Myr, it dramatically fails in reproducing the young stellar clusters. D09 argued that indeed a process creating Ba in the last hundreds Myr of evolution is quite unlikely 
(unless local enrichment is invoked).
In a recent paper, Desidera et al. (2011) presented a thorough investigation of several fundamental properties of the young 
solar-type star HD~61005, a probable member of the Argus association which also includes the IC~2391 OC. 
The authors confirmed the previous findings by D09, obtaining that this star is characterised by extremely high Ba abundance, with [Ba/Fe]=0.63$\pm$0.06 dex.
The origin of this unusual overabundance in Ba is not clear (see D09, M11, and Desidera et al. 2011) and several questions naturally arise from the observed pattern: is the enhancement in the Ba content shared 
by the majority (totality?) of young clusters populating the solar neighbourhood?
Are the nearby young clusters characterised by a unique [Ba/Fe] value? Do they show any intrinsic internal dispersion? 
Do the other $s$-process elements follow the enhancement in Ba?

In an attempt to answer these questions, we present the first $s$-process element (both first-peak Y-Zr,  and second-peak Ba-La-Ce) abundance determination of three young nearby associations:
the AB Doradus, Carina-Near, and Ursa Major moving groups (see \citealt{zukrev} for a review). 
These three associations cover an age range which is particularly critical for our investigation, spanning 
from $\sim$500 Myr (UMa; \citealt{king}), to $\sim$200 Myr (Carina-Near; \citealt{zuk06}), up to $\sim$50$-$110 Myr (AB Doradus; \citealt{zuk04}; \citealt{lu05}; \citealt{messina}).
Moreover, the well-known star $\iota$ Horologii, a probable evaporated member of the Hyades (see Vauclair et al. 2008),
is also included in our analysis. In a forthcoming paper (Biazzo et al. 2012, in preparation) we report iron-peak, $\alpha$-, and odd/even-$Z$ elemental abundances for the same sample stars, 
discussing on more general grounds the chemical abundance pattern of these young stellar aggregates.

The paper is organised as follows:  in Section~\ref{sec:obs} we provide information on our target stars, while the abundance
analysis technique is described in  Section~\ref{sec:analysis}. 
Our results are given in Section~\ref{sec:results} and discussed in detail in Section~\ref{sec:disc}. A summary of our findings
ends the paper (Section~\ref{sec:summary}).

\section{Sample, observations, and data reduction}\label{sec:obs}
The sample was selected from the list of members of
AB Dor, Carina-Near and Ursa Major moving groups, considering
only GK dwarfs with projected rotation velocity $v \sin i \le 15$
km s$^{-1}$ and no evidence of spectroscopic binarity.
High S/N FEROS spectra (R=48000) are available for 11 stars fulfilling
these criteria.
We further added to the sample $\iota$ Hor, a planet host and
probable evaporated member of the Hyades, for which several
abundance analyses based on high-dispersion spectra are available.
The spectra of five stars were acquired as part of the program
for spectroscopic characterization of the targets for the SPHERE
GTO survey (Mouillet et al. 2010), the spectra of $\gamma$ Lep A and B are from
Desidera et al. (2006), while the remaining ones were retrieved from
ESO archive.

Data reduction was performed using a modified version of the FEROS-DRS
pipeline (running under the ESO-MIDAS context FEROS), yielding the
wavelength-calibrated, merged, normalised spectrum through the
following reduction steps: bias subtraction and bad-column masking;
definition of the echelle orders on flat-field frames; subtraction
of the background diffuse light; order extraction; order-by-order
flat-fielding; determination of wavelength-dispersion solution by
means of ThAr calibration lamp exposures; order-by-order normalisation,
re-binning to a linear wavelength-scale with barycentric correction;
and final merging of the echelle orders.
\section{Data analysis}\label{sec:analysis}

\subsection{Abundance measurements}

We determined Y, Zr, Ba, La, and Ce abundances for the Sun and for our sample stars.
Spectral synthesis was applied to all the features under consideration, 
including isotopic shift and/or hyperfine structure (hfs) data as needed.
We employed the MOOG code (Sneden 1973; 2010 version) and the Kurucz (1993) set of model atmospheres, with the overshooting option switched off.
Our analysis relies on a few lines that are proven to be reliable
for this kind of study, namely the $\lambda$4398 \AA~for Y{\sc ii}, the $\lambda$4209 \AA~for Zr~{\sc ii}, the $\lambda$5853 \AA~for Ba~{\sc ii}, the $\lambda$4087\AA~for La~{\sc ii}, and the $\lambda$4073 \AA~for Ce~{\sc ii}. 
We carefully checked in our sample spectra for other suitable lines, but
our search was unfruitful because they are weak, or blended, or both (e.g., the Y~{\sc ii} lines at $\lambda$5087\AA~, $\lambda$5200\AA~, $\lambda$5205\AA, the 
Zr~{\sc ii} ones at $\lambda$4161\AA~and $\lambda$4443 \AA~-see also \citealt{mashonkina07}, 
the $\lambda$4322\AA~and $\lambda$4333\AA~of La~{\sc ii}, the $\lambda$4350 \AA~and $\lambda$4562 \AA~of Ce~{\sc ii}). 
Due to the relatively high-metallicity of our stars, we discarded the Ba~{\sc ii} line at $\lambda$6141 \AA~, which is known to be blended with a strong iron line. 
Moreover, as discussed by \cite{mg00} and \cite{mashonkina07}, the Ba~{\sc ii} line at $\lambda$6496 \AA~can be affected by NLTE effects that can be up to $-$0.2 dex (on average) for stars in the range $-$1.0 $<$[Fe/H] $<$+0.1 dex. 
We focused, hence, on the $\lambda$5853 \AA~feature, which is quite insensitive to NLTE effects
(\citealt{mg00}), as well as strong and isolated.
\begin{center}
\begin{table*}
\caption{Stellar parameters, metallicity, and $s$-process element abundances of our sample stars}\label{t:results}
\begin{tabular}{lccccccccr}
\hline
\hline
Star     	& T$_{\rm eff}$ & log$g$ & $\xi$ & [Fe{\sc ii}/H] & [Y{\sc ii}/Fe] & [Zr{\sc ii}/Fe] & [Ba{\sc ii}/Fe] & [La{\sc ii}/Fe] & [Ce{\sc ii}/Fe]     \\
                &   (K)         &        & (km s$^{-1}$)  &       &                &                &                  &                  &                  \\
\hline
         	&      &     &     &        &                                     &                 &                    &              &      		 \\
	 	&      &     &     &        &               {\bf AB Doradus}     &                 &                    &  		 &     			  \\
	 	&      &     &     &        &                                   &                 &                    &              &       			\\
HIP114530	& 5600 & 4.6 & 1.6 &   0.11 &                     0.00$\pm$0.10 &   0.05$\pm$0.10 &  0.20$\pm$0.15  &   0.00$\pm$0.10    &  0.03$\pm$0.10	 \\
HIP82688 	& 6100 & 4.6 & 1.8 &   0.15 &                       ------	 &    ------	   &  0.27$\pm$0.20  &    ------	  &   -----		 \\
TYC 5901-1109   & 6200 & 4.6 & 1.5 &   0.14 &                     0.04$\pm$0.12 &   0.10$\pm$0.12 &  0.40$\pm$0.15  &  0.00 $\pm$0.07    &  0.06$\pm$0.12	 \\
TYC 9493-838    & 5450 & 4.6 & 1.8 &   0.08 &                     0.04$\pm$0.10 &   0.10$\pm$0.09 &  0.30$\pm$0.15  &  0.00 $\pm$0.07    &  0.06$\pm$0.14	 \\
TYC 5155-1500 	& 5800 & 4.6 & 1.9 &   0.11 &                    $-$0.05$\pm$0.10 &   0.04$\pm$0.10 &  0.10$\pm$0.15  &  0.00 $\pm$0.10    &  0.00$\pm$0.10	 \\
         	&      &     &     &        &                  		 &		   &  		     &		          &	 		 \\
Average  	&      &     &     &        &                      0.00$\pm$0.04   &   0.07$\pm$0.03 &  0.25$\pm$0.11  &  0.00$\pm$0.01	  &  0.04$\pm$0.03	 \\
        	&      &     &     &        &                  		 &		    &  		     &		          &	 			\\
	 	&      &     &     &        &           {\bf Carina Near}       &		    &	             &		          &			 \\
         	&      &     &     &        &                  		&		    &  		     &		          &			 \\
HIP58240 	& 5900 & 4.6 & 1.6 &  0.02  &                 0.00$\pm$0.10	 & 0.07$\pm$0.08   & 0.25$\pm$0.15   &    0.10$\pm$0.10   & 0.03$\pm$0.10  \\
HIP58241 	& 5750 & 4.5 & 1.7 & $-$0.01  &                 0.05$\pm$0.10	 & 0.10$\pm$0.10   & 0.23$\pm$0.18   &    0.00$\pm$0.07   & 0.04$\pm$0.12  \\
HIP37918	& 5550 & 4.7 & 1.7 &  0.09  &                 $-$0.05$\pm$0.10	 & 0.00$\pm$0.10   & 0.20$\pm$0.15   &   $-$0.05$\pm$0.10 & 0.00$\pm$0.10 	   \\
HIP37923 	& 5450 & 4.6 & 1.5 &  0.10  &                 0.00$\pm$0.10	 & 0.12$\pm$0.10   & 0.10$\pm$0.20   &    0.03$\pm$0.07   & 0.04$\pm$0.12  \\
         	&      &     &     &        &                  		&		   &		     &	                  &       		\\
         	&      &     &     &        &                  		&		   &		     &	                  &      		 \\
Average  	&      &     &     &        &                  0.00$\pm$0.05	&  0.07$\pm$0.05   & 0.20$\pm$0.07   &    0.02$\pm$0.06	  &  0.03$\pm$0.02	\\         
         	&      &     &     &        &                  		&		    &  		     &		          &			 \\
	 	&      &     &     &        &        {\bf Ursa Major}        &  	            &		     &		          &  			\\		
$\gamma$LepA 	& 6350 & 4.3 & 1.4 &  0.05 &	                 0.00$\pm$0.10  &   0.05$\pm$0.10  & 0.25$\pm$0.20   &  0.04$\pm$0.10	  &   0.08$\pm$0.10 		\\
$\gamma$LepB 	& 5100 & 4.6 & 1.5 & $-$0.01 &	                 0.00$\pm$0.15  &   0.04$\pm$0.10  & 0.15$\pm$0.15   &  0.02$\pm$0.10	  &	 ----  		\\ 
         	&      &     &     &        &                                 &		  &		     &	                  &     		  \\
         	&      &     &     &        &                                 &		  &		     &	                  &      		 \\
Average  	&      &     &     &        &                  0.00$\pm$0.01     &   0.05$\pm$0.01  &  0.20$\pm$0.07   &  0.03$\pm$0.01	  &	0.08$\pm$0.10			 \\    
         	&      &     &     &        &                  	       &		   &  		     &		          &			 \\
         	&      &     &     &        &               {\bf Hyades (?)}  &		   & 		     &		          &                        \\
$\iota$ Hor 	& 6200 & 4.5 & 1.5 &  0.15  &                 $-$0.05$\pm$0.10	 & 0.06$\pm$0.12   & 0.14$\pm$0.15   &    $-$0.03$\pm$0.10  &  0.00$\pm$0.12 \\
\hline
\hline
\end{tabular}
\end{table*}
\end{center}  
      
There is no information in the literature on the hfs and/or isotopic shifts of Y~{\sc ii}, Zr~{\sc ii}, and Ce~{\sc ii} lines.
However, the main isotope of yttrium is $^{89}$Y, which has an odd mass number, but the level splitting is negligible because of 
the small spin and magnetic momentum of the yttrium nucleus. All four naturally occurring Ce isotopes are even isotopes (nuclear spin I=0), i.e. 
$^{136}$Ce,$^{138}$Ce, $^{140}$Ce, and $^{142}$Ce; due to the lack of hfs splitting and/or isotopic shifts, Ce~{\sc ii} lines are usually narrow.
Of the five Zr isotopes, only one ($^{91}$Zr) has an odd mass number and accounts for about $\sim$11\% of the Zr solar content; however, according to \cite{mashonkina07} 
the hfs effects are negligible.
Thus we adopted a single-line treatment to compute the abundances of Y, Zr, and Ce, taking the log$gf$ values from \cite{hannaford}, \cite{ljung}, and \cite{lawler09}, respectively.

Lanthanum, the first of the rare earth elements, has only one stable isotope ($^{139}$La), and owing to a 
non-zero nuclear spin (I=7/2), it is heavily affected by the hfs splitting.
The log$gf$ values for all its components were retrieved from \cite{lawler01}. 
Finally, although the Ba line at $\lambda$5853$\AA$~does not experience severe hyperfine structure and isotopic shifts, we decided to employ the hfs data
provided by \cite{mcwilliam} to gather robust results\footnote{The hfs treatment has the effect to de-saturate the spectral lines
and, hence, can alter the inferred abundance from barely saturated lines}. We adopted the isotopic solar mixtures of 81\% for ($^{134}$Ba +$^{136}$Ba +$^{138}$Ba) and
19\% for ($^{135}$Ba +$^{137}$Ba). Note that since this line shows negligible hyperfine structure and modest isotopic shifts, the results do not depend on the abundance ratio for even 
and odd isotopes.

Stellar parameters are spectroscopically optimised. T$_{\rm eff}$ values were obtained by zeroing the 
slope of iron abundances from Fe~{\sc i} lines with the excitation potential (typically $\sim$ 100 Fe~{\sc i} lines for each star, covering a wide wavelength range from 4830 \AA~to 7960 \AA~ -see D'Orazi et al. 2009, 2011 for details on the line list atomic parameters). 
Final microturbulence values ($\xi$) were obtained by removing the trend of iron abundances with 
the equivalent widths (EWs), while gravities were derived from the imposition of the ionisation equilibrium condition, i.e. minimising the difference in iron abundances
computed from Fe~{\sc i} and Fe~{\sc ii} lines ($\leq$ 0.05 dex).
The adopted atmospheric parameters are listed in Cols. 2,3,4 of Table~\ref{t:results}; while in Col. 5 we list the [Fe~{\sc ii}/H] abundances used as input metallicity in the model atmosphere 
because all our scrutinised lines are at the first stage of ionisation. 
We note, however, that the agreement between [Fe~{\sc i}/H] and [Fe~{\sc ii}/H] is always within 0.02 dex for all our sample stars.

In order to minimise the impact of the uncertainties on the atomic parameters, we carried out a strictly differential 
analysis with respect to the Sun, optimising the initial line list on a solar spectrum acquired with the same
instrument (at the same resolution) used for our sample stars. 
Adopting for the Sun T$_{{\rm eff}_\odot}$=5770 K, log$g_\odot$=4.44, $\xi_{\odot}$=1.1 km s$^{-1}$, and [Fe/H]$_\odot$=0, we obtained 
the solar abundances listed in Table~\ref{t:sun} (Col.~2). Our results are compared with the 
solar abundances by \cite{grevesse} (Col. 3), with the values by Asplund et al. (2009)
computed using 3D hydrodynamical model atmosphere (Col. 4), as well as with the meteoritic abundances (Col. 5) from CI carbonaceous chondrites taken from  
\cite{lodders}. Within the uncertainties, our values are in agreement with these previous works.

\begin{center}
\setcounter{figure}{0}
\begin{figure*}
\includegraphics[width=16cm]{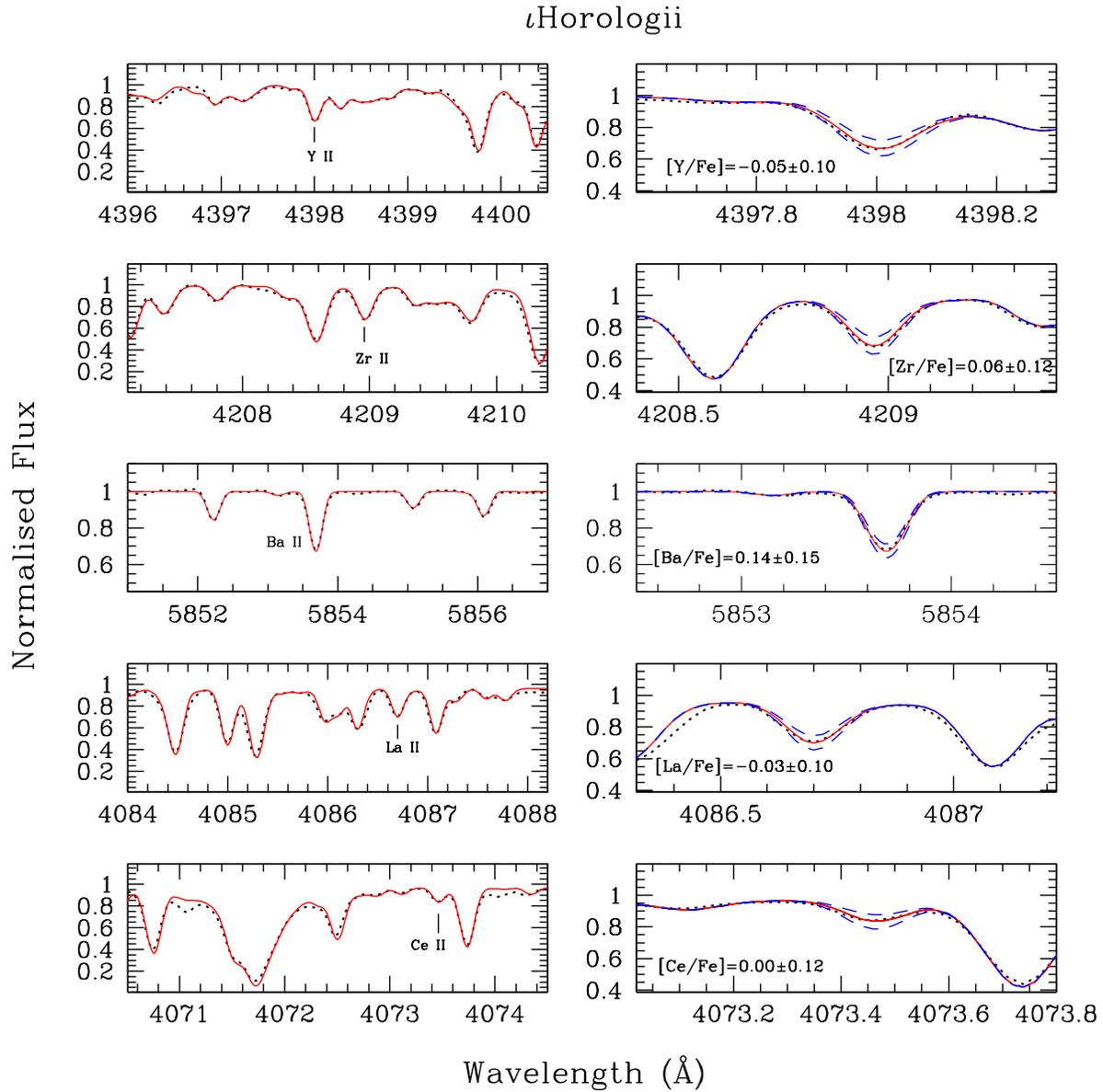}
\caption{Example of spectral synthesis of Y~{\sc ii}, Zr~{\sc ii}, Ba~{\sc ii}, La~{\sc ii}, and Ce~{\sc ii} for the star $\iota$ Horologii. In the left panels we show the wavelength regions around the $s$-process element features, while in right-hand panels we show the best-fit with the corresponding uncertainty.}\label{f:synall}
\end{figure*}
\end{center}
In Figure~\ref{f:synall} we provide an example of spectral synthesis in the wavelength regions surrounding the Y~{\sc ii}, Zr~{\sc ii}, Ba~{\sc ii}, La~{\sc ii}, and Ce~{\sc ii} lines for one
of our sample stars ($\iota$ Horologii); in
the right-hand panels we zoomed on each feature reporting the best-fit value with its corresponding error (see Section~\ref{sec:errors}).
The plot clearly shows that because of the crowding of several relatively strong lines, the continuum placement
is really critical and the synthesis technique is needed to derive reliable abundances.
\begin{center}
\begin{table}
\caption{Solar elemental abundances}\label{t:sun}
\begin{tabular}{lcccr}
\hline
\hline
Element   & This study          &  Grevesse et al.  &	 Asplund et al.      & Meteoritic \\
          &                     &     (1996)        &         (2009)         &         \\
\hline
          &                     &          	    &			     &  	   \\
Y	  & 2.19$\pm$0.05  	&  2.24$\pm$0.03    &	 2.21$\pm$0.05       &    2.17$\pm$0.04     \\
Zr	  & 2.55$\pm$0.04  	&  2.60$\pm$0.03    &	 2.58$\pm$0.04       &    2.53$\pm$0.04     \\
Ba	  & 2.13$\pm$0.05  	&  2.13$\pm$0.05    &	 2.18$\pm$0.09       &    2.18$\pm$0.03     \\
La	  & 1.12$\pm$0.04  	&  1.17$\pm$0.07    &	 1.10$\pm$0.04       &    1.17$\pm$0.02     \\
Ce	  & 1.60$\pm$0.07  	&  1.58$\pm$0.09    &	 1.58$\pm$0.04       &    1.58$\pm$0.02     \\
\hline						 
\hline
\end{tabular}
\end{table}
\end{center}
\subsection{Error estimates}\label{sec:errors}
Two kinds of uncertainties are related to abundances derived from spectral syntheses: $(i)$ the errors owing to the best-fit
determination and $(ii)$ the uncertainties due to the adopted atmospheric parameters.

The first source of uncertainty, which includes also errors in the continuum placement, ranges from 0.07 to 0.14 dex for Y, Zr, La, and Ce. 
Barium has a typical uncertainty of 0.15 dex, reaching a value of 0.2 dex for the fastest rotators (see Table~\ref{t:results}). 

To evaluate the impact of stellar parameters (T$_{\rm eff}$, log$g$, and $\xi$), we varied each quantity separately (leaving the others unchanged) and 
checked the abundance sensitivity to that variation; a change of $\pm$60K in T$_{\rm eff}$, $\pm$0.1 km~s$^{-1}$ in $\xi$, and $\pm$0.1 dex in log$g$
has been adopted as a conservative estimate (see D'Orazi \& Randich 2009; Biazzo et al. 2011; D'Orazi et al. 2011).  
All lines considered in this study are formed by ions dominating the concentration of their corresponding species, 
and are either in the ground-state either in low-excitation levels. This implies that, for a given atmosphere, they are all equally sensitive to 
effective temperature and surface gravity variations (see also \citealt{mashonkina07}).
We found that a change in T$_{\rm eff}$ of $\pm$60K results in a variation of 0.02 dex in [X/Fe] and a change in log$g$ of 0.1 dex leads to an abundance variation of 0.03 dex.
Microturbulence dominates the errors due to stellar parameters: Y, Zr, La, and Ce experience a change of about 0.04 dex, while Ba, which has the strongest feature lying on the 
shoulder of the curve of growth, displays the highest sensitivity, namely $\Delta$[Ba/Fe]=0.07 dex.

The final errors can be obtained by summing in quadrature uncertainties given from these two different contributions, 
i.e., $\sigma_{\rm tot}$=$\sqrt{\sigma_{\rm fit}^2+\sigma_{\Sigma(param)}^2}$. 
Typical uncertainties are of about $\sim$0.10-0.12 dex for Y, Zr, La, and Ce abundances, while Ba is characterised by larger total errors, related to the best fit determination and to the uncertainties on stellar parameters; both aspects reflect the almost saturated behaviour of this strong spectral feature.

\section{Results}\label{sec:results}

Our results are reported in Table~\ref{t:results}, where the [X/Fe] ratios are given for all the elements along with the corresponding uncertainty due to the best-fit
determination (Cols. 6$-$10). For the star HIP82688, a member of the AB Doradus moving group, we could derive only the Ba abundance, 
because the relatively high rotational velocity (vsin$i$=15 km s$^{-1}$, see Table~\ref{t:vsini}) makes the other spectral features blended, hampering the abundance analysis.  
For the first-peak $s$-process elements, in our case Y and Zr, we measured the following average values for the three associations: [Y/Fe]=0.00$\pm$0.04 (rms) dex and [Zr/Fe]=0.07$\pm$0.03 dex for AB Doradus, 
[Y/Fe]=0.00$\pm$0.05 dex and [Zr/Fe]=0.07$\pm$0.05 dex for Carina-Near, [Y/Fe]=0.00$\pm$0.01 dex and [Zr/Fe]=0.05$\pm$0.01 dex for Ursa Major. 
 We found that the three young associations share a homogeneous solar composition as far as the light $s$-process elements (Y, Zr) are concerned 
(the rms for all mean values is significantly lower than the errors).

Focusing on the second-peak elements La and Ce, we obtained: [La/Fe]=0.00$\pm$0.01 dex and [Ce/Fe]=0.04$\pm$0.03 dex for AB Dor, [La/Fe]=0.02$\pm$0.06 dex and [Ce/Fe]=0.03$\pm$0.02 dex
for Carina-Near, and [La/Fe]=0.03$\pm$0.01 dex and [Ce/Fe]=0.08$\pm$0.10 dex for Ursa Major\footnote{Due to the intrinsic weakness of the Ce line, we could not derive the Ce abundances 
for $\gamma$ Lep B (the coldest star of our sample) . 
We indicate as an average value for the cluster the [Ce/Fe] ratio measured in $\gamma$ Lep A quoting, as average error, the uncertainty on the best-fit determination.}. 
The $hs$ process elements La and Ce as well as the first-peak $s$-process elements Y and Zr, show solar ratios.
Similarly, the field star $\iota$ Horologii exhibits solar abundances, with [Y/Fe]=$-$0.05$\pm$0.10 dex,
[Zr/Fe]=0.06$\pm$0.12 dex, [La/Fe]=$-$0.03$\pm$0.10 dex, and [Ce/Fe]=0.00$\pm$0.12 dex (see Table~\ref{t:results}).

On the other hand, the mean barium abundances are [Ba/Fe]=0.25$\pm$0.11 dex for AB Dor, [Ba/Fe]=0.20$\pm$0.07 dex for Carina-Near, and [Ba/Fe]=0.20$\pm$0.07 dex for Ursa Major. 
This means that the three clusters show an enhancement in the [Ba/Fe] ratio of $\sim$0.2 dex, which is not revealed in the other $s$-process elements, neither the first-peak nor second-peak. 
In Figures~\ref{f:teff} and \ref{f:feh} we show the [X/Fe] ratios for all the analysed elements as a function of effective temperature (T$_{\rm eff}$)
and metallicity ([Fe/H]): with the exception of Ba, all the $s$-process element abundances are consistent, within the uncertainties, with a solar pattern.

The [Ba/Fe] seems also to display the largest variation within each association; however, since the scatter is still within the observational errors, we argue that those variations in the measured [Ba/Fe] ratios are simply due to uncertainties in stellar parameters that are dominated by microturbulence. 

For the star $\gamma$ Lep A, \cite{edvardsson93} -hereafter E93- determined Y, Zr, and Ba abundances, finding [Y/Fe]=$-$0.06 dex, [Zr/Fe]=$-$0.03 dex, and [Ba/Fe]=0.05 dex; their results are in very good agreement with our estimates for Y and Zr, while for Ba we obtain higher abundance ($\Delta$([Ba/Fe])=0.2 dex). Differences in effective temperature and 
gravity are negligible, being T$_{\rm eff}$=6398K ($\Delta$(our$-$E93)=$-$48K) and log$g$=4.29 ($\Delta$(our$-$E93)=0.01 dex); on the other hand they derived a notably 
higher microturbulence value, i.e. $\xi$=1.83 km s$^{-1}$. The difference of 0.4 km s$^{-1}$ between our study and that of E93 accounts for the difference 
in the [Ba/Fe] ratio; if we adopt the E93 value of $\xi$=1.83 km s$^{-1}$ then we reproduce their derived value of [Ba/Fe]=0.05 dex. Concerning [Y/Fe] and [Zr/Fe] ratios, we infer values of $-$0.13 and $-$0.10 respectively. These
are consistent with the findings of E93 (within the uncertainties). 
We point out that while the microturbulence value from E93 was
derived from a relationship between $\xi$, log$g$ and T$_{\rm eff}$ (obtained from the
analysis of about 12-17 Fe~{\sc i} lines only), our values are
spectroscopically optimised on more than 100 Fe~{\sc i} lines, covering a wide range in EP, wavelength and strength.
It should be noted that in analysing another Ursa Major member (HR~2047), E93 derived a [Ba/Fe]=+0.25 dex and concluded that their Ba abundance determination might be 
considered normal for young stars in the light of an inverse [Ba/Fe] correlation with age.
Similar conclusions were reached by \cite{castro99}, who determined Ba abundances 
of seven GK members of the Ursa Major moving group; none of their stars are included in our sample, so that a direct comparison of stellar parameters
and abundances can not be performed. However, their results are consistent with our findings, with [Ba/Fe] ratios exhibiting an excess at about $\sim$0.3 dex level.

The enrichment of \textit{only} Barium in relation to the solar s-process pattern is further highlighted in Figure~\ref{f:starsun}. Here, we compare the spectrum of the solar analogue 
(T${\rm eff}$=5750K, log$g$=4.5, and [Fe/H]=$-$0.01 dex), with that of the Sun.  
The latter is convolved to the projected rotational velocity of the former. 
Along with Ba, we focused on the Zr~{\sc ii} and La~{\sc ii}, taking them as representatives of $ls$ and $hs$ elements, respectively.

For HIP58241, we obtained a microturbulence value of $\xi$=1.7 km s$^{-1}$ and hence a difference of +0.6 km s$^{-1}$ compared to the value derived for the sun ($\xi_{\odot}$=1.1 km s$^{-1}$, see Section~\ref{sec:analysis}). 
It can be shown trivially that such a difference in $\xi$  can not account for the difference in EWs of the Ba~{\sc ii} line in the two stars. Given that  EW(Ba~{\sc ii})$_{\odot}=63$ m\AA~and EW(Ba~{\sc ii})$_*= 86$ m\AA~then ($\Delta$(EW)=23 m\AA). If we approximate the spectral line as a rectangle (with side l=EW/2) and keep all the parameters constant, then we can assume that the $\xi$ values are proportional to the EW. Thus, a changing of +0.6 km s$^{-1}$ in $\xi$, results in an enhancement in the EW of about 12 m\AA. 
This implies that in order to reproduce the difference in the EW strength only in terms of microturbulence values, we should assume a value of $\xi$=2.4 km s$^{-1}$ ($\Delta(\xi-\xi_{\odot})$=+1.3 km s$^{-1}$). 
Indeed a spectral synthesis with the latter value of microturbulence yields a [Ba/Fe]=0.00. 
However, such a high value would naturally lead to a significant under-estimate of the iron content (with the condition of no trend between abundances from 
Fe~{\sc i} lines and EW strength no longer satisfied); similarly, abundances for the other $s$-process elements would be shifted towards lower values of about $\sim$0.15 dex.

The comparison shown in Figure~\ref{f:starsun} clearly demonstrates that there is no difference in the strength of the Zr and La features (and hence in the abundances), while the Ba line is significantly stronger in the star.
This means that the high Ba abundance we found for our sample stars is not affected by major systematic uncertainties, reinforcing the reliability of our analysis.
The implications of such a peculiar trend in the Ba content are discussed in Section~\ref{sec:disc}.

As a by-product of the spectral synthesis technique, we derived rotational velocities (v$sin$i) of our sample stars. 
The values we obtain are in very good agreement, within the uncertainties, with previous determinations (see Table~\ref{t:vsini}).
\begin{center}
\setcounter{figure}{1}
\begin{figure*}
\includegraphics[width=14cm]{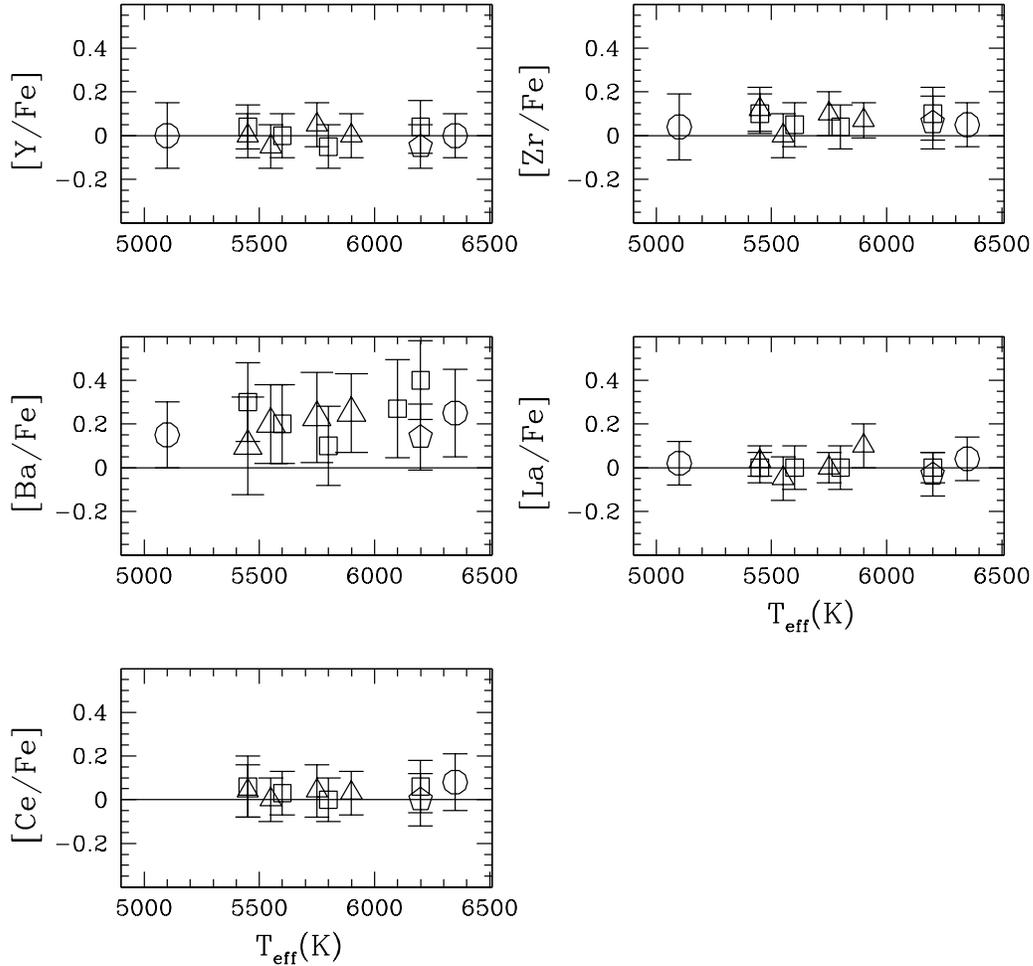}
\caption{[X/Fe] ratios as a function of T$_{\rm eff}$ for all our sample stars. Empty squares, triangles and circles refer to AB Doradus, Carina-Near, and Ursa Major, respectively, while 
$\iota$ Horologii is shown as a pentagonal symbol. Solid line represents the solar value.}\label{f:teff}
\end{figure*}
\end{center}
\begin{center}
\setcounter{figure}{2}
\begin{figure*}
\includegraphics[width=14cm]{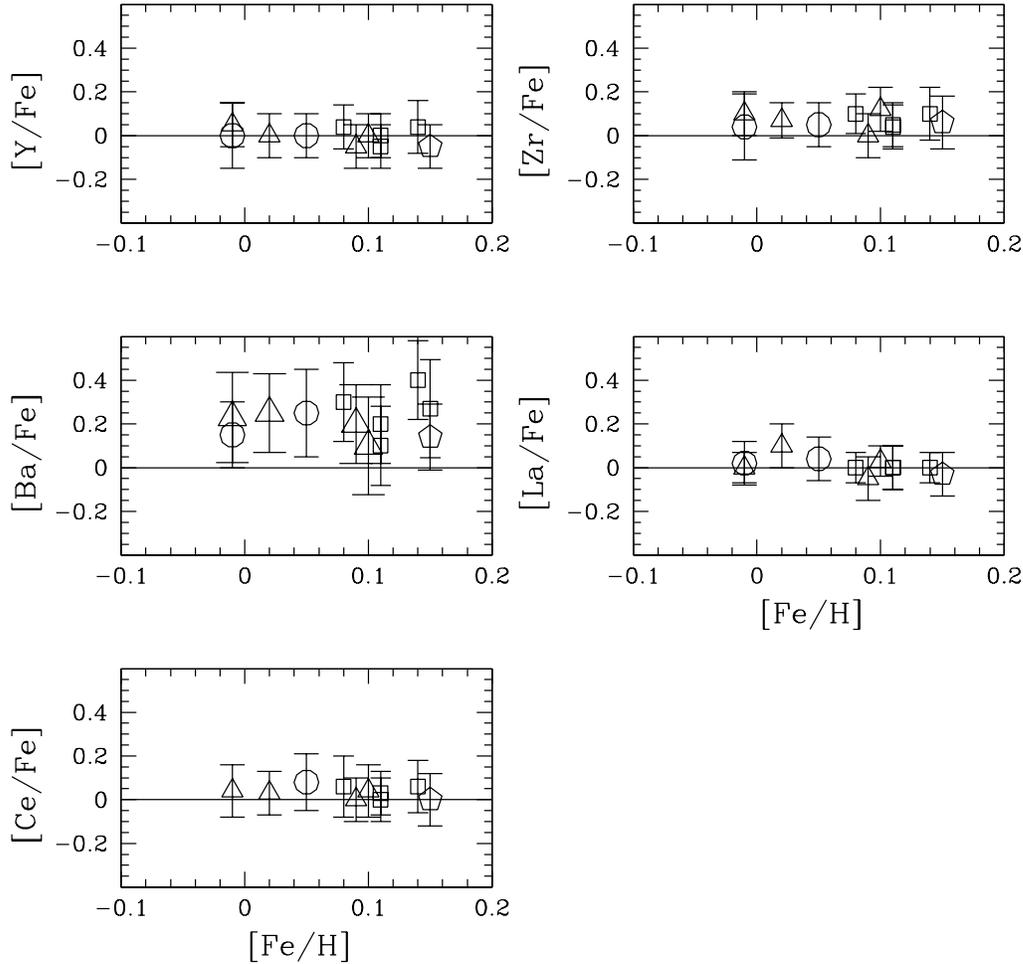}
\caption{[X/Fe] ratios $vs$ metallicity. Symbols as in Figure~\ref{f:teff}}\label{f:feh}
\end{figure*}
\end{center}
\begin{center}
\begin{table}
\caption{Rotational velocities (vsin$i$) from the literature and from our analysis.}\label{t:vsini}
\begin{tabular}{lccr}
\hline
\hline	
Star      &  v$sin$i       &    v$sin$i	  	&           References$^{*}$ \\
          &   (Literature) & (This work)  	&                \\
          &   km s$^{-1}$  & km s$^{-1}$  	&		 \\
\hline
	  &                &              	&  		 \\ 
	  &		   &		  	& 	         \\
HIP114530  &	6.6$\pm$1.2 &    6.0$\pm$2	&	[1]      \\
HIP82688   &    15.0$\pm$1.5    &   15.0$\pm$2	&	[2]         \\
TYC 5901-1109   &	6.0$\pm$2.0 &    5.0$\pm$2	&	[3]         \\
TYC 9493-838    &	3.0$\pm$0.1 &    3.0$\pm$2      &       [1] 	 \\
TYC 5155-1500  &     9.0$\pm$2.0 &    8.0$\pm$2	&	[3] 	 \\
HIP58240  &	5.2$\pm$1.2 &	5.0$\pm$2	&	[1] 	 \\
HIP58241  &	9.0$\pm$1.2 &	9.0$\pm$2	&	[1] 	 \\
HIP37918  &	6.3$\pm$1.2 &	6.0$\pm$2	&	[1] 	 \\
HIP37923  &	 3.2$\pm$1.2 &	3.0$\pm$2	&	[1] 	 \\
$\gamma$LepA &	7.7$\pm$1.8    &	8.0$\pm$2	&	[4] 	 \\
$\gamma$LepB &	2.8$\pm$1.8   &	3.0$\pm$2	&	[4] 	 \\
$\iota$ Hor  &	6.2$\pm$1.2 &	6.0$\pm$2	&	[1] 	 \\
\hline
\hline
\end{tabular}
\begin{list}{}{}
\begin{footnotesize}
\item[$^\mathrm{*}$] [1] Torres et al. (2006); [2] \cite{zuk04}; [3] \cite{dasilva09}; [4] \cite{silvano06}
\end{footnotesize}
\end{list}
\end{table}
\end{center}   
\begin{center}
\setcounter{figure}{3}
\begin{figure}
\includegraphics[width=8cm]{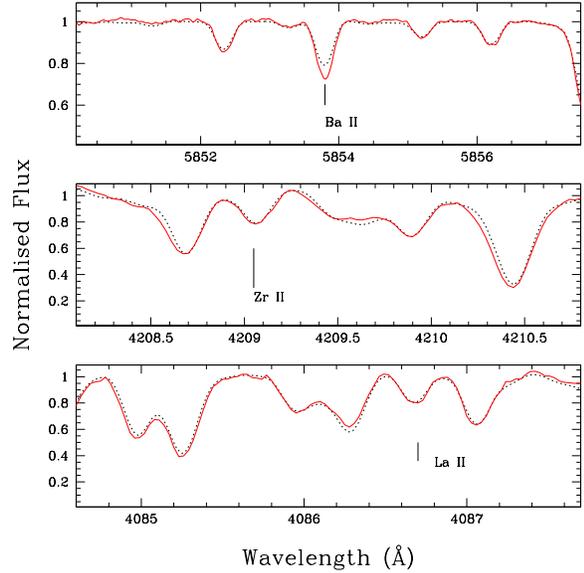}
\caption{Comparison between the spectrum of the star HIP58241 (solid line) and the solar one (dotted line) convolved with the rotation profile 
at the target vsin$i$.}\label{f:starsun}
\end{figure}
\end{center}

\section{Discussion}\label{sec:disc}
In this section we discuss the scientific implications of the derived abundances. Our observations suggest that  
the three young associations have a solar $s$-process abundance pattern concerning Y, Zr, La, and Ce, while Ba seems to be considerably over-abundant. 
In Section~\ref{sec:sproc} we discuss the global $s$-process element pattern observed in young open clusters and associations, while  
because of its outstanding nature, Ba is discussed separately in Section~\ref{sec:barium}.
\subsection{The $s$-process content of young stellar clusters}\label{sec:sproc}
\begin{center}
\setcounter{figure}{4}
\begin{figure*}
\includegraphics[width=14cm]{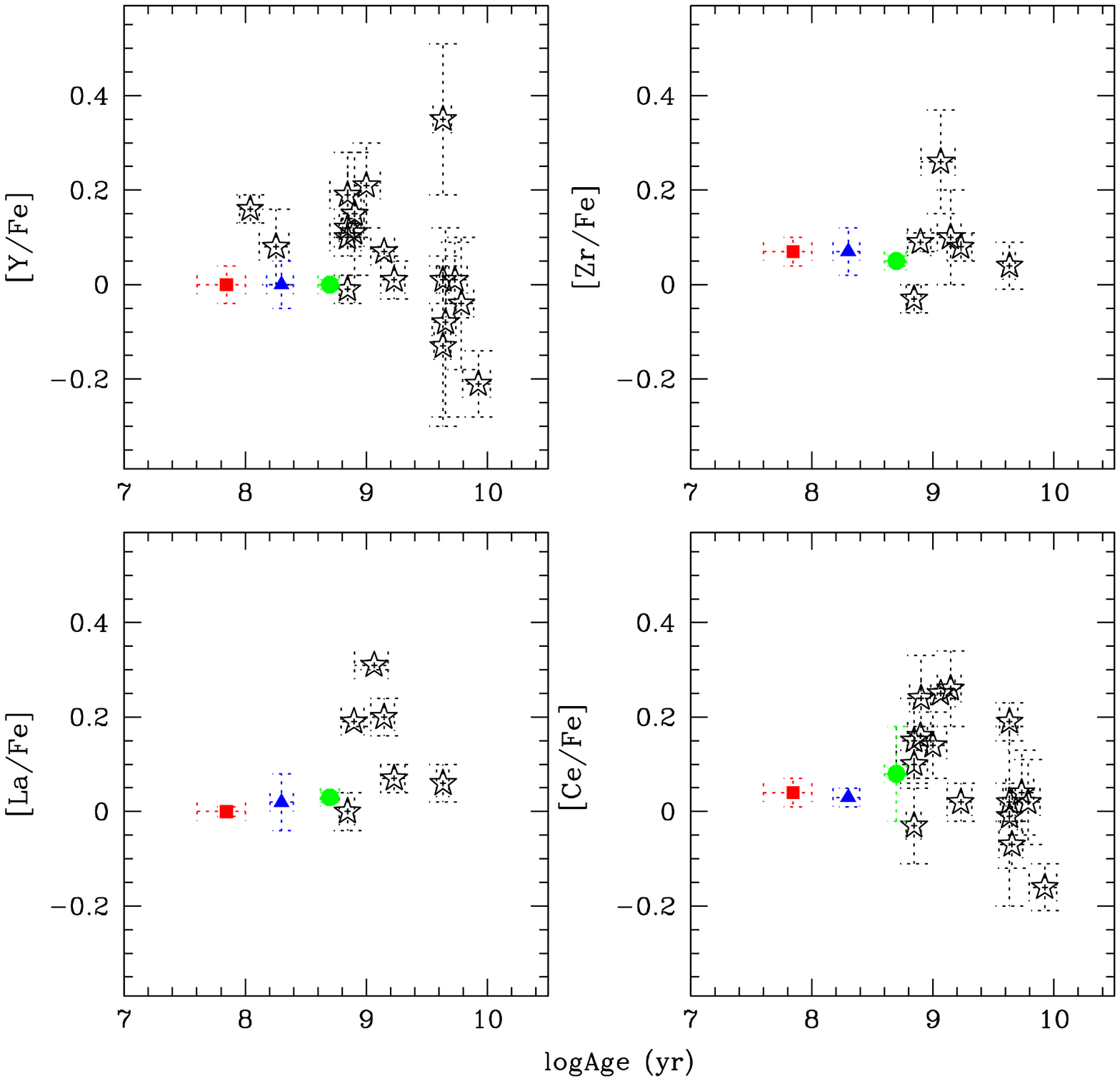}
\caption{[X/Fe] ratios as a function of age for OCs analysed by M11 (starred symbols) and for our sample clusters (AB Dor, Carina-Near, and Ursa Major). Symbols are as for Figures~\ref{f:teff} and~\ref{f:feh}.}\label{f:ageall}
\end{figure*}
\end{center}
Recently, M11 derived abundances of Y, Zr, La, and Ce for 19 OCs which complemented the Ba determinations of D09. 
M11 suggested that younger clusters are characterised by a higher $s$-process element content (both first-peak and second-peak $s$-process elements) than older ones. The authors claimed that a steep increase is seen in OCs with ages between 1.5 and 0.5 Gyr, after which 
the [X/Fe] values remain constant around $\sim$0.2 dex. They concluded that such a trend is best explained if low-mass AGB stars (1$-$1.5 M$_\odot$) are more effective in the production
of $s$-process elements than previously thought (see M11 for details on the anti-correlation between the AGB mass and the efficiency in the 
$n$-capture nucleosynthesis). 

Moreover, M11 highlighted that at any given age a scatter in the heavy-element content is present, suggesting that it might be partly related to uncertainties in the [Fe/H] determinations, 
but that an intrinsic (real) variation cannot be excluded. Currently, a theoretical explanation for such a scattered trend is not in hand.

In Figure~\ref{f:ageall} we show the [X/Fe] ratios derived by M11 (starred symbols) as a function of age (see Magrini et al. 2009), along with abundances for the 
three associations studied here.
Firstly, we note that we can not gather reliable indication on trends of Zr and La with cluster age because, for those elements, only six clusters were analysed by M11; consequently we focus our discussion only on Ce and Y. 
In fact, by considering only the M11 targets, the linear correlation coefficient $r$ for Zr and La as a function of age is $-$0.16 and $-$0.29, respectively, and is not statistically meaningful. Conversely, 
[Y/Fe] and [Ce/Fe] ratios show an anti-correlation with the cluster age at more than 99\%  significance level, being $r_Y$=$-$0.59  (16 degrees of freedom) and $r_{Ce}-$=0.66 (14 degree of freedom).
Note that [Ce/Fe] measurements for the two youngest clusters of the M11's sample are not available. 
If we omit Berkeley 29 (an outlier cluster, see the discussion in M11) from our calculations, then we find
$r_Y$=$-$0.82 and $r_{Ce}$=$-$0.59.
We also computed the Spearman rank correlation coefficient (which is less affected from the presence of outliers) and found $r_{\rm Spear}$=$-$0.57 for Y ($-$0.68 without Be 29) 
and $r_{\rm Spear}$=$-$0.41 for Ce ($-$0.45 without Be29).

The inclusion of our three young associations is of particular relevance in this context allowing us to enlarge the sample in the critical range from $\sim$50 $-$ 500 Myr. 
It is worth emphasising that despite the uncertainty affecting the age determination of these young moving groups\footnote{The age of AB Doradus is still under discussion, ranging from 50 to 110 Myr; 
for our purpose we adopt an average nominal age of $\sim$ 70 Myr \citep{zuk11}.}, the central point of our discussion is the relative difference in age. 
Thus, we can confidently assert that AB Doradus and Carina-Near are significantly younger than the Hyades OC, and slightly older than IC~2602/IC2391 ($\sim$35-50 Myr), while the Ursa Major moving group is marginally younger than the Hyades.

If we consider our inferred abundances along with those of M11, 
two possible scenarios can be deduced from Figure~\ref{f:ageall}:

$(i)$ There is a systematic offset between our study and M11. This may be due, for instance, to the different adopted techniques (i.e., EWs $vs$ spectral synthesis; see Section~\ref{sec:analysis}). In this sense a re-analysis of their sample clusters by means of spectral synthesis would be necessary to verify this point (see also discussion in the following Section).

$(ii$) Alternatively, we detected a true variation (at roughly 0.2 dex level) in the $s$-process abundances of young nearby stellar populations, co-existing both clusters with solar and over-solar heavy-element abundances.
Under this assumption, our findings confirm that a scatter in $s$-process elements exists for young, open clusters agreeing with the spread found in older clusters by M11. We recall again that there is no current explanation for such a spread (see also M11). 
However, in order to definitely disentangle the source of this variation we need to enlarge the sample of $s$-process element abundances in young OCs:
a homogeneous, extensive observational survey on 
clusters younger than the Hyades is essential to draw final conclusions and to provide a strong observational evidence and constrain (perhaps new) theoretical models.

\subsection {The Barium issue}\label{sec:barium}

In Figure~\ref{f:ageba} we plot [Ba/Fe] as a function of the age  for our associations and those OCs previously analysed by D09\footnote{Because a systematic offset between clusters analysed using dwarfs and giants
emerged from the D09 sample, we focus only on clusters for which the analysis is based on dwarf stars (see D09).}: 
within the uncertainties (typical values are $\sim$ 0.15 dex), the three young associations fit fairly well with the global trend defined by the other OCs of comparable ages. 
The slightly lower [Ba/Fe] ratios we found for AB Dor and Carina-Near with respect to their coeval clusters analysed in D09 (NGC~2516 and NGC~6475) 
probably reflect the fact that the microturbulence values are on average
larger for our sample stars. 
Further evidence on the peculiar nature of the Ba abundance comes from the Hyades candidate-member $\iota$ Horologii:
for this star we inferred a [Ba/Fe]=0.14$\pm$0.15 dex, which is only slightly smaller than the average value found by D09 for the Hyades OC (i.e. [Ba/Fe]=0.30$\pm$0.05): given the uncertainties, we can conclude that the two abundances are in agreement. 
 
An analogous value for the Ba content of the Hyades was obtained by
\cite{carrera}, who measured [Ba/Fe]=0.36$\pm$0.04 dex. 
Interestingly, they confirm the remarkable nature of the Ba abundances: all the other $s$-process elements exhibit solar ratios, with [Y/Fe]=$-$0.09$\pm$0.03 and [La/Fe]=$-$0.08$\pm$0.04
(note that they found almost identical ratios for the coeval Praesepe cluster). 
Similar outcomes were presented by \cite{gds06}, who confirmed that while Ba is enhanced ([Ba/Fe]=0.35$\pm$0.07), the other heavy element abundances 
result in [Zr/Fe]=$-$0.05$\pm$0.06, 
[La/Fe]=$-$0.14$\pm$0.07, and [Ce/Fe]=0.06$\pm$0.07 (i.e., solar, within the errors).

The comparison of our results with previous studies confirms the Ba (and only Ba) over-abundance; M11 proving to be the only exception. They found super-solar abundances for other elements in the Hyades (e.g., [X/Fe]=0.12$\pm$0.04 for Y and Ce).
Because we are comparing studies that make use of different techniques, models and codes, we are left with large scope for uncertainty. We do not draw any conclusions on the detected abundance pattern; we only emphasise in this context the importance of homogeneous abundance determinations, carried out by means of spectral synthesis, for large sample of Hyades members, including both dwarfs and giants.

\begin{center}
\setcounter{figure}{5}
\begin{figure}
\includegraphics[width=8cm]{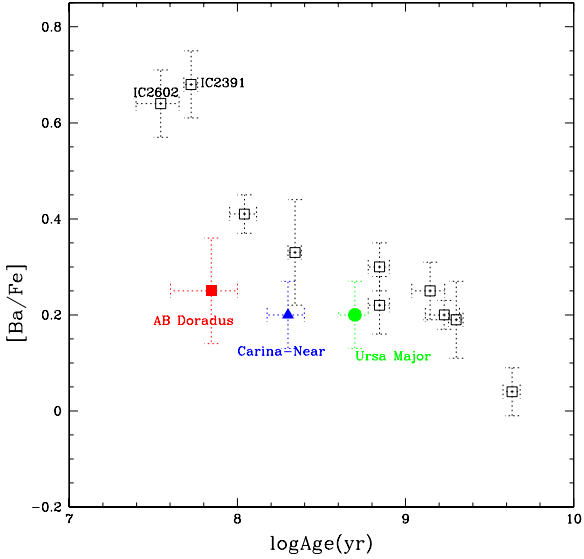}
\caption{Run of Ba abundances with age for our three associations and for the OCs analysed by D09.}\label{f:ageba}
\end{figure}
\end{center}
Most intriguing is the fact that none of the currently available theoretical models can 
account for an enhancement in Ba, without bearing analogous (over-) production of the other $s$-process elements, especially La and Ce (\citealt{busso99}; \citealt{tr04}).
This has led us to suggest that the super-solar Ba abundances are probably related to some (different?) conspiring effects. 

Firstly, the high Ba abundance might be related to the uncertainties in stellar parameters, especially the ones pertaining to the microturbulence values
(see Section~\ref{sec:errors}). At the moment, there is no consensus in the literature whether young, active stars are characterised by higher microturbulence values due to the presence of strong magnetic fields
(see \citealt{steen}; James et al. 2006; Santos et al. 2008; D'Orazi \& Randich 2009; Biazzo et al. 2011). 
However, we demonstrated in Section~\ref{sec:results} that the higher $\xi$ value (in comparison to the Sun) inferred for one of our sample stars cannot account for the difference in the EW strength. The EW is significantly larger in the star implying an over-solar Ba abundance.
In addition, in our analysis there is no obvious trend between the microturbulences (ranging from 1.4 to 1.9 km s$^{-1}$) and the derived [Ba/Fe] ratios; 
the same was found by D'Orazi et al. (2009) for the pre-main sequence clusters IC~2602 and IC~2391, where the Ba enhancement reaches levels of $\sim$0.6 dex.
As a consequence, although the microturbulence may have a significant impact on the inferred Ba abundances, it cannot be an exhaustive explanation.

As further investigation, we checked if  
the unusually high Ba abundance is related to the chromospheric activity level. 
At solar metallicity, the effective depth of the $\lambda$5853 \AA~ Ba~{\sc ii} line formation is at $\log \tau_{5000}$ = $-$1.9, while formation is at  $\log \tau_{5000}$ = $-$3.3 and $\log \tau_{5000}$ = $-$4.8 for Ba~{\sc ii} lines at 6141\AA~and 4554 \AA, respectively. 
Given that the line formation zone is quite deep in the atmosphere, we should not expect large impact from the above hot chromospheres. As information on chromospheric activity is available for stars in our sample we can directly verify this prediction and look for any correlation between Ba overabundance and (i) CaII H\&K chromospheric emission, (ii) coronal emission
and (iii) rotational velocity. The logR$_{\rm HK}$ values were measured on the same spectra used for
the abundance analysis as in Desidera et al. (2011), while for the components
of $\gamma$ Lep logR$_{\rm HK}$ was taken from Desidera et al. 2006.
X-ray luminosity for the targets was obtained looking for X-ray counterparts
of our targets in the ROSAT All Sky Surveys
(Voges et al. 1999;2000); calibration to X-ray fluxes was performed
following Hunsch et al. (1999).
Individual X-ray luminosities for the components of $\gamma$ Lep are
from \cite{schmitt04}.
Projected rotational velocities are from Table 3. 
As expected, we found that there is no relationship between these activity indicators and the Ba over-abundance.
We note also that Ba is mostly ionised at the temperatures 
of our sample stars; therefore any over-ionising effect 
is expected to be almost negligible (see Schuler et al. 2003, 2004; D'Orazi \& Randich 2009 for over-ionisation/excitation effects on e.g., Fe, Ti, Ca in young open clusters).

An ``indirect" chromosphere-related effect could be the shape of the stratification in temperature as a function of the optical depth, namely T($\tau$), in the model atmosphere. 
Due to the presence of a hot chromosphere, one would expect a T($\tau$) function less steep compared to that of old stars (the outer atmosphere should be heated at a certain extent by the upper chromosphere levels).
To verify if this so we modified our model atmospheres code, manually varying the T($\tau$). We found that the effect is negligible in the opposite direction (i.e., log$\epsilon$(Ba) is marginally higher).
This is not surprising, because as explained above, the Ba~{\sc ii} line at 5853 \AA~ forms relatively deep in the stellar atmosphere. 
As a consequence, we can discard this explanation.

Finally, NLTE barium abundances were also derived using the fitting between observed and 
calculated profiles of the $\lambda$5853 \AA~ line. The NLTE profile of this line was 
generated using the updated version of the MULTI code (Carlsson 1986; Korotin et al. 1999), 
and the barium atomic model that is described in details in Andrievsky et al. (2009). 
For a correct comparison of the NLTE barium line profile with the observed one, 
we used a combination of the NLTE (MULTI) and LTE synthetic spectrum code. 
The LTE synthetic spectrum was calculated for a given wavelength range that 
includes 5853 \AA~ line. For the barium line, the corresponding $b$-factors 
(which express the deviation from LTE level populations), that were calculated with 
MULTI, are included in the LTE synthetic spectrum code, where they are used in the 
calculation of the barium line source function. The lines of other species in 
vicinity of the barium line are synthesized using the input atomic data from 
VALD\footnote{http://ams.astro.univie.ac.at/vald/}. After comparison of NLTE and LTE 
barium abundance derived from the $\lambda$~5853 \AA~ line, we reached the conclusion 
that NLTE effect is negligible for this line (about 0.02 dex).

We do not possess a straightforward answer to the peculiar trend of the Ba abundance in young clusters/associations and several possibilities (including indirect effects related to the chromospheric activity level) remain open.
The further enhancement at even younger ages ($\lesssim$50 Myr), reaching $\sim$0.6 dex, 
deserves special attention, invoking urgent observational surveys for the determination 
of Ba (and other $s$-process elements) for large samples of stars in very young clusters (age range 10-50 Myr).
We plan to attend to this point in the near future.
\section{Summary and conclusions}\label{sec:summary}

In this paper we presented $s$-process Y, Zr, Ba, La, Ce abundance determinations in three nearby young associations, i.e. AB Doradus, Carina-Near, and Ursa Major.
Our results can be summarised as follows:
\begin{itemize}
\item[$-$] The three associations show solar abundance ratios for both first-peak and second-peak elements, here Y, Zr, La, and Ce.		

\item[$-$] The [Ba/Fe] is notably enhanced in all our sample stars, with average value around $\sim$0.2 dex.

\item[$-$] None of the currently available theoretical models can reproduce such enhancement, accounting for a Ba production without a simultaneous enrichment in the other $s$-process elements, especially the second-peak ones (La and Ce).

\item[$-$] We investigated whether the relatively high chromospheric activity of these young stars might play a role in the high Ba abundances we derived. 
However, we do not find the evidence of any obvious correlation between the [Ba/Fe] ratio and activity 
indicators. Moreover, uncertainties due to the abundance analysis 
(including stellar parameters), possible effects of the stratification in temperature of the model atmosphere, and NLTE corrections can be excluded as exhaustive explanations. 
We stress, however, that other possible chromosphere-related effects can be at work and, at this stage, we cannot yet 
totally discard this scenario. 

\end{itemize} 

We conclude stressing that our work reinforces the need for a large, homogeneous investigation of $s$-process abundances in clusters younger than the Hyades to 
draw final conclusions on this issue and provide observational constraints to new theoretical models.

\section*{Acknowledgments}
This paper makes use of data collected for the preparation
of the SPHERE GTO survey.
We warmly thank the SPHERE Consortium for making them available
for the present work.
Based on observations made with the European Southern Observatory
telescopes (program IDs: 70.D-0081(A), 082.A-9007(A), 083.A-9011(B), 
084.A-9011(B)) and data obtained from the ESO Science Archive Facility under request
numbers:143106, 143382, 147882, 152598, 153529, 162614, 6572, 10960.
This work made use of SIMBAD (operated at CDS, Strasbourg, France), NASA Astrophysics Data System Bibliographic Services, and of the WEBDA 
open cluster database, currently maintained by E. Paunzen and C. St\"{u}tz (http://obswww.unige.ch/webda). 
KB acknowledges the financial support from the INAF Post-doctoral fellowship.
VD thanks G. Angelou for his helpful comments.
The anonymous referee is kindly acknowledged for a careful reading of the paper and her/his valuable suggestions which improved the quality of the manuscript.

\label{lastpage}

\end{document}